\documentclass[11pt]{article}
\usepackage{amsmath,amsfonts,amssymb}
\usepackage{latexsym}
\usepackage{dcolumn}
\usepackage{graphicx,epsfig}
\usepackage{amsthm}
\usepackage{color}
\usepackage{graphicx}

\begin{document}
\setcounter{page}{1}

\pagestyle{plain} \vspace{1cm}
\begin{center}
\Large{\bf The role of an invariant IR cutoff in late time cosmological dynamics}\\
\small \vspace{1cm} {\bf K. Nozari$^{a,b,}$\footnote{knozari@umz.ac.ir(Corresponding Author)}}\quad and \quad {\bf P. Dehghani$^{a,}$\footnote{p.dehghani@stu.umz.ac.ir}} \\
\vspace{0.25cm}
$^{a}$Department of Physics, Faculty of Basic Sciences,\\
University of Mazandaran,\\
P. O. Box 47416-95447, Babolsar, Iran\\

\vspace {0.35cm}

$^{b}$Research Institute for Astronomy and Astrophysics of Maragha
(RIAAM)\\ P. O. Box 55134-441, Maragha, Iran
\end{center}

\vspace{1cm}
\begin{abstract}
We study the role of an invariant infra-red cutoff in the late time cosmological dynamics.
This low energy cutoff originates from the existence of a minimal measurable uncertainty
as a result of the curvature of background manifold, and can be encoded in an extended uncertainty relation.
Inspired by black hole entropy-area relation, we extend the analysis to the
thermodynamics of apparent horizon of the universe. By treating both the Newtonian
and general relativistic cosmologies, we show that the contribution of infra-red cutoff in the equations of dynamics can be interpreted as an
effective fluid which is capable of explaining late time cosmic speed up and even transition to a phantom phase of expansion.
We use the latest observational data from PLANCK2018 to constrain the parameter of an extended uncertainty relation.
\end{abstract}

\section{Introduction}
Ordinary quantum mechanics suffers from the lack of precision when it comes to consider the scale at which
gravitation is not negligible. Different sub-leading quantum gravity theories such as string theories, noncommutative geometry,
loop quantum gravity and the doubly special relativity theories have revealed this shortcoming of ordinary quantum mechanics~\cite{Hossenfelder2013}.
Generally, the relevant results would be model-dependent. But a simple \textit{Gedanken Experiment} shows that, within a very small length scale, gravity comes into act and impedes a limit on the further resolution of the adjacent (nearby) spacetime points. To distinguish two spatially nearby points of spacetime, a limit on length resolution must be considered. Gravity imposes a limit on the resolution of adjacent points which lies within the Planck length scale. In result, gravity obstructs to distinguish two points of space-time separated by the length less than the Planck length. In consequence, gravity directly affects the resolution of space-time points. This feature is encoded in a generalized uncertainty principle which forms a minimal uncertainty in position measurement~\cite{Hossenfelder2013}, resulting , for instance, a \emph{micro-black hole}~\cite{Scardigli1999}. Scale at which gravitation
meaningfully affects ordinary quantum mechanics has come to be known as the Planck scale~\cite{Adler2010}. Consequently, in the realm of high-energy
physics, a modified version of Heisenberg Uncertainty Principle (HUP) is established based on ultra-violet cutoff of Planck
energy, which is called Generalized (Gravitational) Uncertainty Principle (GUP). To be more concrete, GUP is not the most general form of quantum
uncertainty principle as the symmetry of symplectic geometry necessitates the possibility of having the form of HUP at which minimal momentum
would be also plausible. In fact, canonical commutation relation $[x, p]=i\hbar$ is founded upon Poisson bracket of $x$ and $p$ of a classical system within a phase space in which its geometry can be handled using closed 2-form $\{dx \wedge dp\}$ resulting a sympelctic structure. Within a symplectic geometry, the notion of area (volume) in phase space is considered instead of length or distance. Then $x$ and $p$ must be viewed in the same footing and symmetrically within a symplectic manifold. Therefore, it is reasonable to assume the symmetry of $x$ and $p$ in terms of 2-forms within a symplectic manifold, representing a physical system which results two modified Heisenberg commutation relations, one containing $p^{2}$ (that is, $[x , p] = i\hbar(1+\beta p^{2})$) and the other containing $x^{2}$ (that is, $[x , p] = i\hbar(1+\alpha x^{2})$) . In view of this argument, both the minimal length and minimal momentum would be plausible taking the symmetry of symplectic geometry into account. For more details, we refer to ~\cite{Nozari2016} where the authors proved, for the first time, that natural cutoffs are due to compactness of symplectic manifolds. When considering the behavior of a quantum particle within a curved background geometry, a minimal amount of momentum uncertainty comes into play. This minimal uncertainty nontrivially results a minimal measurable momentum which plays the role of an IR cutoff. In the language of uncertainty relations, this modification is established as an Extended Uncertainty Principle (EUP)~\cite{Hinrichsen1996,Bolen2005}.

The modified quantum mechanics, based on the above considerations, can lead to a modified version of black hole horizon entropy.
As Bekenstein proposed that black holes do have entropy and information, the second law of thermodynamics was generalized to contain a black hole system.
This generalization is founded upon the notion that black holes must possess information. Bekenstein incorporated rules of thermodynamics and the laws of black hole mechanics to reach the result~\cite{Bekenstein1973}. More elaborately, we do know that information
comprises microstates, resulting the impact of quantum mechanics and its modifications to the entropy of a quantum black hole. Indeed, second law of thermodynamics would be violated without considering the fact that black holes must possess information and then entropy. As an example, suppose an arbitrary object, e.g. a cup, with some information and entropy, is dropped into a stationary black hole. For an external observer who cannot find the cup, its information is lost and the second law of thermodynamics would be violated in view of this observer. For solving this problem, Bekenstein conjectured that black holes must have information. Then the decrease of information within the exterior region of black hole is compensated by the increase of information of the black hole. In result, generalized second law of thermodynamics was proposed by Bekenstein in which the information and entropy never decrease. Bekenstein utilized the generalized second law of thermodynamics, and the laws of black hole mechanics proposed by Bardeen \emph{et al.} (1973), to acquire the entropy of a stationary black hole. But entropy of a black hole is grasped in terms of microstates which are governed by quantum mechanics. Therefore, it is reasonable to expect the impact of quantum mechanics on the black hole entropy as microstates, that make up information of a black hole, behave in the language of quantum mechanics.

Although GUP-corrected black hole entropy has been studied widely so far, less attention has been paid for its EUP counterpart\footnote{see~\cite{Gorji2016} for an attempt in this direction.}, demanding much more studies. The EUP brings a natural cutoff as a minimal measurable momentum, leading to an infra-red regulator. Our attention here is devoted to probe the role of a non-Euclidean quantum mechanics in the correction of black hole entropy, and then to generalize the result to the apparent horizon of the universe. In this manner, we are going to see the effect of this natural infra-red cutoff on the late time dynamics of the universe. In fact, it is interesting to see how an IR cutoff, initiating from quantum mechanical considerations in a curved background, can affect large scale dynamics of the universe. In other words, while an IR cutoff essentially originates from quantum mechanical considerations in a curved background, its possible role in large scale physics is important to be treated carefully, the main goal of this study. To do this end, we adopt the entropic force framework of Verlinde~\cite{Verlinde2011} which states that gravity is essentially an emergent phenomenon with an entropic origin. We treat the black hole thermodynamics in the framework of the extended uncertainty relation to find an IR-modified entropy-area relation. Then we extend the analysis to the apparent horizon of the universe. By focusing on the cosmological dynamics, in both the Newtonian and general relativistic cases, we show that the effect of IR cutoff can be interpreted as an effective fluid with an equation of state parameter that crosses the phantom divide line\footnote{We note that the latest observational data released by the Planck Collaboration~\cite{PLANCK2018} shows that the current value of the equation of state parameter of the cosmic fluid (an effective fluid governing the late time cosmic dynamics) is given by  $w = -1.03\pm0.03$. Then, the equation of state parameter of the cosmic fluid is capable of crossing the cosmological line $w= -1$ and possibly makes a transition to the phantom phase. Therefore, if a model has the potential to provide a new alternative for cosmic fluid, this effective fluid should explain transition to the phantom phase in a natural way. As we shall see, this is actually the case in our model.}, and is capable of explaining the late time cosmic speed up. Then we use the latest observational data from PLANCK2018 in order to constrain the EUP parameter. This study sheds light on the role of an IR cutoff (which normally seems to be irrelevant to quantum mechanical considerations) in treating the large scale dynamics of the universe.

\section{EUP-corrected entropy of AdS black hole}

As of numerous works which signal the possibility of a minimal measurable length within the quantum scale
in which gravitation is effective, we can consider the quantum behavior of physical systems within low-energy regime taking
the background geometry of space to be non-Euclidean. In this case, some works have already been done showing the modification of
HUP when one considers a constant curvature geometry~\cite{Kempf1995,Mignemi2010}(see also~\cite{Nozari2012}).
Previous works have explicitly manifested the modification of HUP by an order proportional to the cosmological constant.
If Anti de Sitter/de Sitter were regarded as the background geometry~\cite{Sokolowski2016} (as observations are
in accord with de Sitter space-time), pertinent modified HUP would be of the following form

\begin{equation}
\delta x\delta p \geq \frac{\hbar}{2}\Big(1+\alpha \Lambda (\delta x)^{2}\Big),
\end{equation}
where $\Lambda = \pm\frac{3}{R^{2}}$ and $R$ is the AdS/dS curvature radius. This form of HUP features minimal amount of energy and also
momentum of a physical system which physically originates from the background curvature of space-time in essence. The problem we are intended to address is
to consider this extended version of HUP to reach a new modified relation for entropy of a Schwarzschild-AdS black hole horizon. We
then generalize this relation to the entropy of apparent horizon of the universe in order to see how the holographic principle acts at this scale, and how an IR cutoff affects the cosmological dynamics. In this regard, Bekenstein entropy is modified by manipulating the impact of curved background geometry on the quantum
behavior of the system. The incorporation of these cutoffs, in general, and for the case of IR cutoff specifically, reduces the number of microstates in the phase space of the system, resulting a reduction of entropy for such a system.

To commence the task, a quantum particle of energy $E$ and size $d$ is considered to be distant from the event
horizon by the order of its Compton wavelength \footnote{We consider an infinitesimally small particle of size $d$ to be distant from the event horizon of an AdS-Schwarzschild black hole by the order of Compton wavelength of the particle following Verlinde~\cite{Verlinde2011}. Since wave-particle duality admits to have a particle of definite size behaving as being a quantum wave packet and then featuring uncertainty, our premise would be right. Although particle features a compact support, it can represent spatial uncertainty as behaving like a quantum wave with, for instance, a Gaussian distribution. The way of thought in this part resembles the approach of Christodoulou~\cite{Christodoulou1970} (see also Medved \emph{et al.}~\cite{Medved2004}).}. As the particle merges into the black hole horizon, amount of the information carried by the particle would add to the horizon. Then entropy of the black hole would be subject to alter slightly~\cite{Medved2004}. By inverting the relation (1) to find $(\delta p)_{min}$, the minimum energy of the test
particle can be acquired
\begin{equation}
 \delta p \geq \frac{\hbar}{2}\Big((\delta x)^{-1}+\alpha \Lambda \delta x\Big) \Rightarrow (\delta p)_{min} = \frac{\hbar}{2}\Big((\delta
 x)^{-1}+\alpha \Lambda \delta x\Big) \Rightarrow E_{min} = \frac{\hbar}{2}\Big((\delta x)^{-1}+\alpha \Lambda \delta x\Big)\,.
\end{equation}
Also it is persuasive to consider the size of the particle as its position uncertainty, that is,
\begin{equation}
d \sim \delta x\,.
\end{equation}
Previous studies have derived the minimal area change of a Schwarzschild black hole due to merging of a particle into the event
horizon in the context of general relativity as~\cite{Christodoulou1970},

\begin{equation}
(\Delta A)_{min} = 8 \pi l_{p}^{2} Ed\,.	
\end{equation}
Now the quantum behavior of the system should be incorporated, so that $8 \pi$ is replaced by a factor $\lambda$. It is important to note that $\lambda$ is a factor yet to be determined via further progress in quantum gravity (see \cite{Medved2004} for a detailed discussion in this regard). The minimal change of entropy happens by adding just one bit of information, $(\Delta S)_{min} = b$. Setting $\hbar = 1$, one finds
\begin{equation}
 (\Delta A)_{min} = \lambda l_{p}^{2} Ed = \frac{\lambda l_{p}^{2}}{2}\Big((\delta x)^{-1}+\alpha \Lambda \delta x\Big)\delta x\,,
\end{equation}
so
\begin{equation}
 \frac{(\Delta S)_{min}}{(\Delta A)_{min}} \simeq \frac{2b}{\lambda l_{p}^{2} \Big(1+\alpha \Lambda (\delta x)^{2}\Big)}\,.
 \end{equation}
By expanding the right hand side of this relation in Taylor series and keeping the terms up to the second order of $\alpha$, we find
\begin{equation}
\frac{(\Delta S)_{min}}{(\Delta A)_{min}} \simeq \frac{2b}{\lambda l_{p}^{2}} \Big(1-\alpha \Lambda (\delta x)^{2} +
\alpha^{2} \Lambda ^{2} (\delta x)^{4}\Big)\,.
\end{equation}

For a moment we consider the content energy of the black hole to be conserved, so that black hole is characterized by a
microcanonical ensemble. To proceed further, outward energy of the black hole relevant to Hawking radiation must be the same
as the inward energy of the merging particle. Establishing this correspondence, the Compton wavelength of the particle results in as follows
\begin{equation}
\frac{1}{2}kT_{H} \sim \frac{1}{\lambda_{c}} \Rightarrow \lambda_{c} \sim \frac{1}{\kappa}\,,
\end{equation}
where $\kappa$ is the surface gravity of the black hole. Now, the task is reduced to obtain the surface gravity of a Schwarzschild-AdS black hole which has been given in Ref.~\cite{Socolovsky2018} as
\begin{equation}
\kappa = \frac{M}{r_{h}^{2}}+\frac{r_{h}}{R^{2}}\,,
\end{equation}
where $M=M(r_{h},R)$. Therefore, one gets
\begin{equation}
M=\frac{r_{h}}{2}\Big(1+\frac{r_{h}^{2}}{R^{2}}\Big)\Rightarrow \kappa = \frac{1}{2r_{h}}+\frac{3r_{h}}{2R^{2}}\,,
\end{equation}
where $r_{h}$ is the radius of the event horizon. With this result for $\kappa$, equation (8) gives $\lambda_{c}$ as follows
\begin{equation}
\delta x \sim \lambda_{c} \sim \frac{1}{\frac{1}{2r_{h}}+\frac{3r_{h}}{2R^{2}}}\,.
\end{equation}
Now by rewriting equation (7), we find
\begin{equation}
\frac{dS}{dA} \simeq \frac{2b}{\lambda l_{p}^{2}}\bigg(1-\frac{\alpha \Lambda R^{4} A}{(3A+4 \pi R^{2})^{2}} + \frac{\alpha^{2}
\Lambda^{2} R^{8} A^{2}}{(3A+4 \pi R^{2})^{4}}\bigg)\,.
\end{equation}
By integrating this result and keeping the terms up to the second order of the deformation parameter, we find
\begin{align}
S(A)\approx\frac{A}{4  l_{p}^{2}}+\frac{\alpha R^{2}}{12\,  l_{p}^{2}} \ln (3A+4 \pi R^{2})& +\frac{\alpha(1-\frac{\alpha}{27}) R^{4}}{4  l_{p}^{2}(3A+4 \pi
R^{2})}\\ \notag
&+\frac{\pi\alpha^{2}}{3 l_{p}^{2}}\frac{R^{6}}{(3A+4 \pi R^{2})^{2}}-\frac{4\pi^{2}\alpha^{2}}{9l_{p}^{2}}\frac{R^{8}}{(3A+4 \pi R^{2})^{3}}\,.
\end{align}
Note that we have set $\frac{2b}{\lambda}=\frac{1}{4}$ in order to reach a result for the entropy-area relation that recovers the Bekenstein-Hawking standard relation $S(A)=\frac{A}{4  l_{p}^{2}}$ in the non-deformed case \footnote{Note that, based on the argument presented in Ref.~\cite{Medved2004}, $8\pi$ can be replaced by $\lambda$ to imply the quantum behavior of the physical system. But $\lambda$ is not an arbitrary constant, yet having a value probably in the order of unity. Therefore, we cannot make $(\Delta S)_{min}$ arbitrarily large by choosing arbitrarily large $\lambda$. Note also that the IR-deformed entropy-area relation presented in this work is in consent with the one obtained from other approaches, like loop quantum gravity, as mentioned in the relation (14).}. This deformed entropy-area relation, which has been obtained in the context of an EUP with a natural IR cutoff, has some interesting features and outcomes as follows:\\
\begin{itemize}
	\item It reduces to the standard Bekenstein-Hawking entropy-area relation in the absence of the minimal momentum (IR) cutoff.
	\item It gives a leading, logarithmic correction term in accordance with the corresponding result obtained in string theory, loop quantum gravity and also noncommutative geometry. In fact, these approaches give a result as
\begin{equation}
S(A)=\frac{A}{4l_{p}^{2}}+c_{0}\ln\Big(\frac{A}{4l_{p}^{2}}\Big)+\sum_{n=1}^{\infty}C_{n}\Big(\frac{A}{4l_{p}^{2}}\Big)^{-n}+const.
\end{equation}
It is obvious that our result is consistent with this general expression, specially for the case $A=4\pi R^{2}$. We note that, as argued in Ref.~\cite{Medved2005a,Medved2005b}, the sign of the logarithmic prefactor may depend on the type of ensemble one adopts. Also the existence of this logarithmic correction terms is dependent on the dimensionality of spacetime in a fascinating manner~\cite{Nozari2007}.
	\item It gives a negative logarithmic prefactor (note that $\alpha$ is a negative quantity for AdS background in this setup).
    \item It gives the exact value of prefactor as $\frac{\alpha R^{2}}{12  l_{p}^{2}}$ which has a trivial geometric interpretation: it is related to the AdS/dS curvature radius $R$ and therefore to the cosmological constant as $\Lambda = \pm\frac{3}{R^{2}}$.
    \item The functional dependence of subsequent correction terms in this relation to the horizon area are in accordance with the corresponding result obtained in other approaches, that is, dependence on $A^{-1}$.
    \item The case $A=4\pi R^{2}$ considers the possibility of equality of black hole event horizon area with the area of apparent horizon. This is an interesting case that will be studied in detail in the next section.

\end{itemize}

\section{Generalization to apparent horizon}

The retrieved entropy-area relation in the last section can be generalized to apparent horizon of the universe
contemplating in mind that all horizons do obey the same rules of thermodynamics, so that holographic principle is invariantly
working for all local and global horizons. By considering the apparent horizon to be bound to the 3-space of $t = const.$
cross-section within the neck of the background topology, the above relation can be more elaborately neat as follows
\begin{align}
A_{horizon} = 4 \pi R^{2} \Rightarrow S(A) = \frac{A}{4l_{p}^{2}}\bigg[1+\underbrace{4 \alpha \Big(\frac{1}{256
\pi^{2}}-\frac{\alpha}{1000}\Big)+\frac{\alpha}{12}\ln (4A)}_\text{EUP correction}\bigg]\,.
\end{align}
Since $\alpha$ is negative for the AdS background, the overall modification to the Bekenstein entropy, up to the second order in EUP parameter, would be
negative. Then the amount of information on the horizon is diminished due to the quantization of microstates, aside from
the background geometry. Moreover, impact of the size of the universe on this relation is explicitly weakened by the
logarithmic factor. In this regard the outcome is in affinity to that of GUP-corrected entropy-area relation (14), but different in some sense as not featuring inverse
area terms.

\section{Impact of IR cutoff on standard cosmology}

Now we are going to see the role of an IR cutoff on cosmological dynamics. Our basic input is the entropy-area relation obtained in the presence of an IR cutoff.
To proceed, we adopt the Verlinde's entropic picture of forces.
Verlinde showed the emergence of Newtonian gravity and also inertial forces by putting entropic picture into the work~\cite{Verlinde2011}. As the constantly working engine of the universe, the second law of thermodynamics is meant to be responsible for gravity, space-time and also expansion of the
universe. This revolutionary idea caused the entropic cosmology to become widespread in redundant works in recent years. There are
somehow different methods in this regard both dependent to (see for instant~\cite{Santos2012}) and independent of the test particle's character, therefore the outcome
would feature either the failure of equivalence principle or not. Here, by incorporating the newly obtained modified
entropy-area relation, we choose the method that is independent of the test particle's character. Then we manipulate the
modified entropy to reach the target that is retrieval of the modified Friedmann equation and the pertinent impact on cosmological
aspects, specially late time dynamics of the universe. We note that the issue of cosmological dynamics within
the entropic picture has been treated by some authors~\cite{Cai2010,Sheykhi2010}. Here our main goal is to see the impact of an IR cutoff on entropic formalism of the cosmological dynamics.

\textbf{Recovering Newtonian cosmology}: We consider a test particle of mass $m$ in the vicinity of the apparent horizon as
distant as proportional to its Compton wavelength. As the particle merges into the horizon, one bit of information would add to the horizon, then entropy of the universe would be magnified by a least factor. Additionally, universe encompasses mass $M$ with
rest energy of $Mc^{2}$, and the holographic principle tells us to seek for information on the horizon consisting of $N$ bits of
information with equipartition energy of $\frac{1}{2}kT$ each. Lastly, as similar to Verlinde's ansatz, we consider area of the
apparent horizon to be proportional to the number of bits of information by an unknown factor.

As contemplating the universe to be a microcanonical ensemble, the first law of thermodynamics leads us to,
\begin{align}
F\Delta x = T\Delta S\,,
\end{align}
where $\Delta x = \eta \lambda_{c} = \frac{\eta}{mc}$ and $T$ is given via the relation $E = \frac{1}{2} NkT$. By setting $N = \frac{A}{Q}$ where $Q$ is a quantity to be determined, we get $\Delta A = Q \Delta N$. Since $\Delta S = \frac{\partial
S}{\partial A} \Delta A$, we find $\Delta A = Q$\, for one bit of information. Then, by using entropy-area relation as given by (15), we obtain the following relation for the entropic force with a natural IR cutoff

\begin{align}
F = \frac{MmQ^{2} c^{3}}{8 \pi k \eta l_{p}^{2} R^{2}} \bigg[1 + 4 \alpha \Big(\frac{1}{256 \pi^{2}} - \frac{\alpha}{1000}\Big) +
\frac{\alpha}{12} \Big(\ln (4A) + 1\Big)\bigg]\,.
\end{align}
By setting $\frac{Q^{2} c^{3}}{8 \pi k \eta l_{p}^{2}} \equiv G$\,, we reach at
\begin{align}
 Q = \sqrt{8 \pi k \eta}\hspace{0.05cm}	 l_{p}^{2}\,,
\end{align}
and therefore
\begin{align}
 N = \frac{A}{\sqrt{8 \pi k \eta} \,\,l_{p}^{2}}\,,
\end{align}
which is in accordance with the Verlinde's proposal. Finally, the modified Newtonian gravitational force by taking an EUP, that admits a natural IR cutoff into account, is given by the following relation
\begin{align}
F = G \frac{Mm}{R^{2}} \bigg[1 + 4 \alpha \Big(\frac{1}{256 \pi^{2}} + \frac{1}{48}- \frac{\alpha}{1000}\Big) + \frac{\alpha}{12} \ln (16 \pi
R^{2})\bigg]\,.
\end{align}
Now by setting $R = a(t)r$, where $a(t)$ and $r$ are the cosmic scale factor and a proper distance respectively, the modified Newtonian cosmology comes out in this setup,
\begin{align}
\frac{\ddot{a}}{a} = - \frac{4 \pi G}{3} \rho \bigg[1 + 4 \alpha \Big(\frac{1}{256 \pi^{2}} + \frac{1}{48}- \frac{\alpha}{1000}\Big) +
\frac{\alpha}{12} \ln (16 \pi R^{2})\bigg]
\end{align}
where $\rho = \frac{M}{V} $ is the density of the universe. We compare this result with its counterpart in standard model of cosmology
\begin{equation}
\frac{\ddot{a}}{a} = - \frac{4 \pi G}{3} \rho(1+3w)\,
\end{equation}
where $w=\frac{p}{\rho}$ is the equation of state parameter with $p$ being the pressure of the cosmic fluid. Comparing the last two equations we can deduce
\begin{equation}
w=\frac{4}{3} \alpha \Big(\frac{1}{256 \pi^{2}} + \frac{1}{48}- \frac{\alpha}{1000}\Big) +
\frac{\alpha}{36} \ln (16 \pi R^{2})\,.
\end{equation}
Recall that Eq. (21) is obtained through an entropic argument in the presence of a natural IR cutoff, that is encoded by $\alpha$ in the theory.
We argue that the effect of IR cutoff is revealed as an effective fluid which its equation of state parameter depends on $\alpha$ and the area of the apparent horizon of the universe, $A=4\pi R^{2}$. In other words, in this comparison, we are able to introduce an effective fluid with equation of state parameter as given by Eq. (23). Figure 1 shows the behavior of equation of state parameter versus $\alpha$, where $\alpha$ is a negative parameter in this setup.

\begin{figure}[ht!]
\centering
\includegraphics[width=12cm]{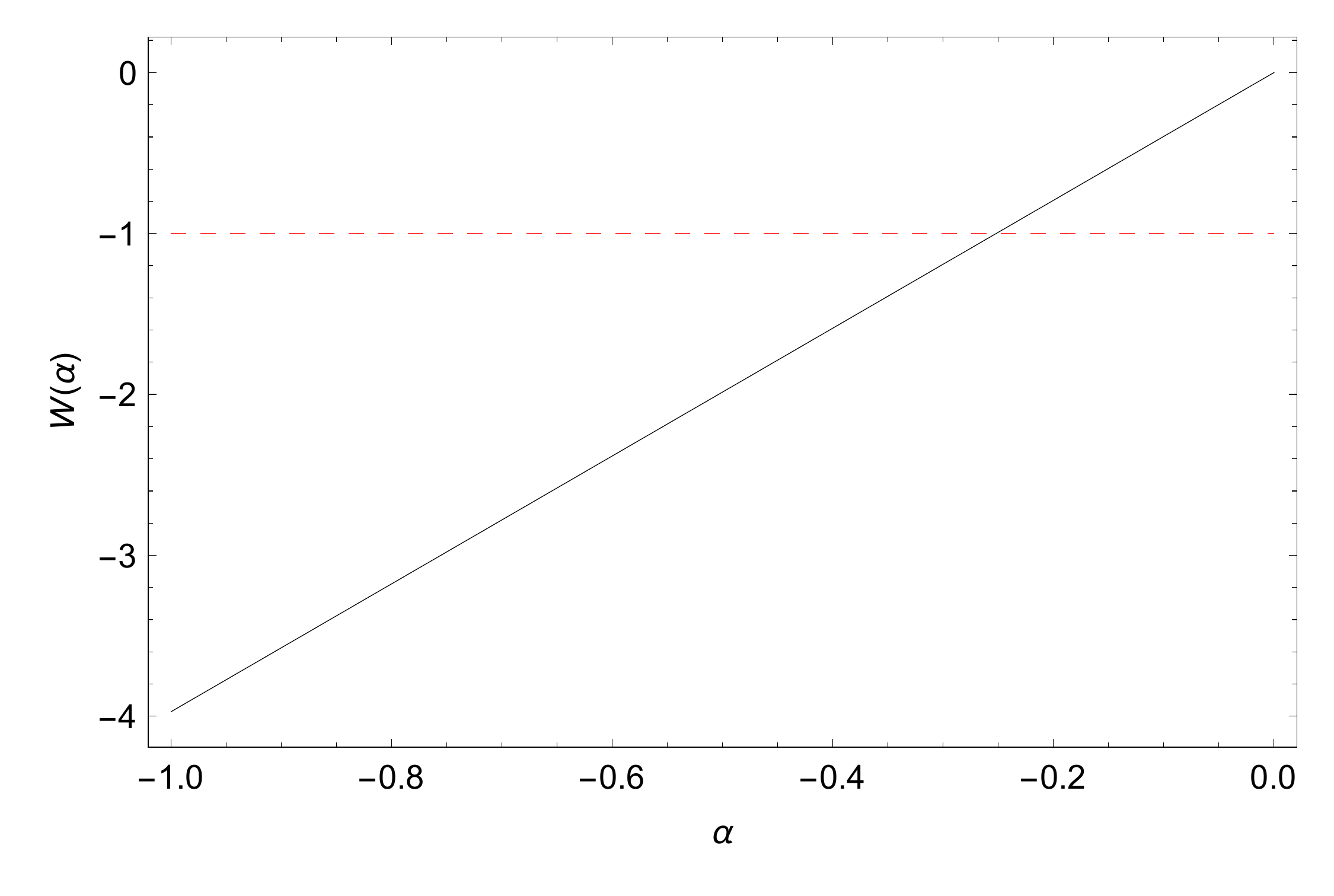}
\caption{Equation of state parameter of the effective fluid versus $\alpha$. The size of the observable universe has been set to $R=10^{30}$m. \label{Fig1}}
\end{figure}


This figure shows that the introduced effective fluid can play the role of a dark energy component. The equation of state parameter of this effective fluid evolves, by increasing the absolute value of $\alpha$, from a quintessence-like region ($-1<w(\alpha)<-\frac{1}{3}$), then crossing the cosmological constant line at $\alpha=-\,0.252$, and transits to the phantom-like region. Therefore, the effective fluid defined here provides a reliable cosmic history at the late time. The recent data from PLANCK satellite combined with type Ia supernovae (SNe) for the effective equation of state parameter of the cosmic fluid gives
$w =-1.03\pm0.03$ \cite{PLANCK2018}. This observational data provides us with a constraint on the EUP parameter $\alpha$ as $-1.06\,\leq\,w\,\leq -1.00$ which gives
\begin{equation}
-1.06\,\leq\,\bigg[\frac{4}{3} \alpha \Big(\frac{1}{256 \pi^{2}} + \frac{1}{48}- \frac{\alpha}{1000}\Big) +
\frac{\alpha}{36} \ln (16 \pi R^{2})\bigg]\,\leq -1.00\,,
\end{equation}
that, with $R=10^{30}$m, finally gives $-\,0.267\,\leq\,\alpha\,\leq -\,0.252$\,. \\

\textbf{Relativistic Cosmology:} The integral idea how to move from Newtonian cosmology to the relativistic one is to regard
mass of the universe as what is realistically causing the universe expand. Previous works have shown that the active gravitational mass
contributing to the expanding universe is formulated by Komar mass~\cite{Wald1984},
\begin{align}
M_{_{Komar}} = 2\int \Big(T_{\mu \nu} - \frac{1}{2} T g_{\mu \nu}\big)u^{\mu}u^{\nu}dV\,.
\end{align}
Calculation for a perfect fluid energy-momentum tensor as seen by a comoving observer leads to the density of the universe replaced by $(\rho + 3p)$.
In result the acceleration equation takes the following form
\begin{align}
 \frac{\ddot{a}}{a} = - \frac{4 \pi G}{3} \Big(\rho + 3p\Big) \bigg[1 + 4 \alpha \Big(\frac{1}{256 \pi^{2}} + \frac{1}{48}-
 \frac{\alpha}{1000}\Big) + \frac{\alpha}{12} \ln \Big(16 \pi R^{2}\Big)\bigg]\,,
\end{align}
which can be rewritten in the following compact form
\begin{align}
\frac{\ddot{a}}{a} = - \frac{4 \pi G}{3} \Big(\rho + 3p\Big) \Big(1 + B + C \ln  R\Big)\,,
\end{align}
where $B=B(\alpha)$ and $C=C(\alpha)$. Consequently the overall modification would be of logarithmic character. If $\alpha$, as a negative parameter, is considered to be of the order of the value obtained in the previous part ($\alpha\sim-\,0.26$, see after Eq. (24)) and $R \sim 10^{30}$m, the quantity inside the second parenthesis in the right hand side would be negative, therefore $\ddot{a}$ would be positive, and the universe undergoes a positively accelerating expansion. In result, this model is capable of explaining positively accelerating expansion of the universe. In other words, there are some subspaces of the model parameter space in the favor of a positively accelerated expansion of the universe as late time cosmic speed up in our IR playground. To proceed further, equation (26) has to be multiplied by $a \dot{a}$, moreover continuity equation has to be used for $(\rho + 3p)$ term. After some calculation we arrive at the following relation
\begin{align}
\dot{a}^{2}+k = \frac{8 \pi G}{3} \bigg\{\int d(\rho a^{2}) \Big[1 + 4 \alpha \Big(\frac{1}{256 \pi^{2}} + \frac{1}{48}- \frac{\alpha}{1000}\Big)\Big] +
\int \ln  (16 \pi R^{2})d(\rho a^{2})\bigg\}\,.
\end{align}
 Now by using $p = w \rho$ where $w$ is a constant and $\rho = \rho_{0} a^{-3(1+w)}$ for adiabatic expansion ($\rho_{0}$ is the present time density of the universe), the last integral can be generally solved. After all, the modified Friedmann equation is achieved as follows
\begin{align}
H^{2} + \frac{k}{a^{2}} = \frac{8 \pi G}{3} \rho \bigg[1+ \underbrace{\frac{\alpha}{6(1+3w)} + 4 \alpha \Big(\frac{1}{256 \pi^{2}} +
\frac{1}{48}- \frac{\alpha}{1000}\Big) + \frac{\alpha}{12} \ln (16 \pi R^{2})}_ \text{EUP correction} \bigg]\,,
\end{align}
which can be written in the following compact form
\begin{align}
H^{2} + \frac{k}{a^{2}} = \frac{8 \pi G}{3} \rho \Big(1+ D + E \ln R \Big)\,,
\end{align}
where $D=D(\alpha)$ and $E=E(\alpha)$. Two important issues should be addressed here:
\begin{itemize}
\item
Firstly one can define an effective energy density as
\begin{equation}
\rho^{(eff)}= \rho\Big[1+\frac{\alpha}{6(1+3w)} + 4 \alpha \Big(\frac{1}{256 \pi^{2}} +
\frac{1}{48}- \frac{\alpha}{1000}\Big) + \frac{\alpha}{12} \ln (16 \pi R^{2}) \Big].
\end{equation}
Since $\rho$ is positive definite by definition, the quantity inside the bracket should be positive in order to have a positive definite $\rho^{(eff)}$. Owing to the negativity of $\alpha$, this is possible for equation of state parameter, $w$, in the vicinity of $w=-\frac{1}{3}$ where $w<-\frac{1}{3}$, that is for $w\rightarrow (-\frac{1}{3})^{(-)}$. So, there are some subspaces in the model parameter space that ensure the positivity of the effective energy density. As $w \rightarrow (-\frac{1}{3})^{(-)}$, the energy density of the universe takes the positive definite value independent of $\alpha$ ($\alpha$ is negative). But for $w=-1$ that is supportive for accelerating expansion, $\alpha$ must be greater than $-0.08$ for positive definite energy density of the universe. This means that $-0.084 < \alpha < 0$. Since the fluid with equation of state parameter $w$ is supposed to obey required energy condition, with the same argument, the effective fluid is also capable to respect the same energy conditions. Note that, as we have seen in the previous part, the observational constraint on effective equation of state parameter gives the condition $-\,0.267\,\leq\,\alpha\,\leq -\,0.252$. At the same time, the positive definiteness of effective energy density requires $-0.084 < \alpha < 0$. Therefore, if we demand the effective fluid to resemble a quintessence-like feature, then the acceptable range of parameter $\alpha$ is $-0.084 < \alpha < 0$. But if we let the effective fluid, in addition to the quintessence-like behavior, be able to mimic a phantom-like behavior, then the range $-\,0.267\,\leq\,\alpha\,\leq -\,0.252$ is acceptable. \\
\item
Secondly, for a spatially flat spacetime, $k=0$, by writing the Friedmann equation (29) as
\begin{equation}
\Big(H^{(eff)}\Big)^{2}=\frac{8 \pi G}{3} \rho\,,
\end{equation}
where
\begin{equation}
\Big(H^{(eff)}\Big)^{2}=\frac{H^{2}}{f(\alpha)}\,,
\end{equation}
with
$$f(\alpha)=1+\frac{\alpha}{6(1+3w)} + 4 \alpha \Big(\frac{1}{256 \pi^{2}} +
\frac{1}{48}- \frac{\alpha}{1000}\Big) + \frac{\alpha}{12} \ln (16 \pi R^{2}),$$
eq. (32) transforms the effect of IR cutoff into the geometric part of the Einstein's equation (Friedmann equation). Since $f(\alpha)$ can be greater than unity, at least in some subspaces of the model parameter space, the result gives a reduced Lemaitre-Hubble parameter. Then, the effect of an IR cutoff in the late time cosmological dynamics can be translated into a purely geometrical language in this fashion.
\end{itemize}

\section{Summary and Conclusion}

While a minimal uncertainty in position measurement due to the finite resolution of spacetime points addresses the existence of a natural ultra-violet cutoff as a minimal length, a finite resolution in momentum space requires a minimal measurable momentum as an infra-red cutoff. This minimal measurable momentum originates from curvature of the background manifold, and plays the role of an infra-red regulator for quantum field theory. Incorporating these natural cutoffs in the standard quantum mechanics via generalization of the uncertainty relations, or deformation of the Heisenberg algebra, reveals a variety of phenomenological aspects of the final quantum gravity proposal. In this paper we studied the role of an infra-red cutoff, which is essentially a low energy effect, on the cosmological dynamics of the universe at the late time. In fact our purpose was to see how an IR cutoff, that initiates from quantum mechanical considerations in a curved background, can affect large scale dynamics of the universe. To achieve this goal, our strategy was to study thermodynamics of event horizon of an AdS-Schwarzschild black hole in the presence of an IR cutoff. We have derived a modified entropy-area relation which has several novel and interesting results. It contains a leading logarithmic correction term in accordance with other well-known approaches to quantum gravity such as the loop quantum gravity. The main achievement here is that we were able to find an exact geometric logarithmic prefactor which is related to the AdS/dS curvature radius $R$ and therefore to the cosmological constant as $\Lambda = \pm\frac{3}{R^{2}}$. In order to treat the role of IR cutoff on cosmological dynamics, we extend our analysis to the thermodynamics of apparent horizon of the universe relying on the Verlinde's entropic gravity conjecture. By focusing on the thermodynamics of apparent horizon, we were able to find IR-deformed Friedmann and acceleration equations. By working in both the Newtonian and general relativistic cases, we have shown that the effect of IR cutoff can be interpreted as an effective fluid with an equation of state parameter that crosses the phantom divide line and is capable of explaining the late time cosmic speed up. In fact, incorporation of the IR cutoff shows itself as an effective fluid whose equation of state parameter mimics an observationally viable dark energy component at the late time. In comparison with the latest observational data from PLANCK2018, we were able to constrain the parameter of the IR-deformed uncertainty relation. Finally we emphasize that an IR cutoff, which normally seems to be irrelevant to quantum mechanical considerations, plays a crucial role in exploring the large scale dynamics of the universe at the late time.\\

{\bf Acknowledgement}\\
The work of K. Nozari has been supported financially by Research
Institute for Astronomy and Astrophysics of Maragha (RIAAM) under
research project number 1/6025-48.

\end{document}